\documentclass[aps,prd,twocolumn,showpacs,amsmath,amssymb]{revtex4-1}
\usepackage{amsmath}
\usepackage{graphicx}
\usepackage{subfigure}
\usepackage{epstopdf}
\usepackage{color}
\usepackage{multirow}
\usepackage{setspace}
\usepackage{overpic}
\usepackage{amssymb}
\usepackage{braket}
\usepackage{makecell}
\usepackage[bookmarksnumbered, pdfstartview=FitH,colorlinks,urlcolor=blue, citecolor=blue,linkcolor=blue] {hyperref}
\usepackage{lineno}
\usepackage{bm}
\usepackage{rotating}
\usepackage[utf8]{inputenc}

\newcommand{\BESIIIorcid}[1]{\href{https://orcid.org/#1}{\hspace*{0.1em}\raisebox{-0.45ex}{\includegraphics[width=1em]{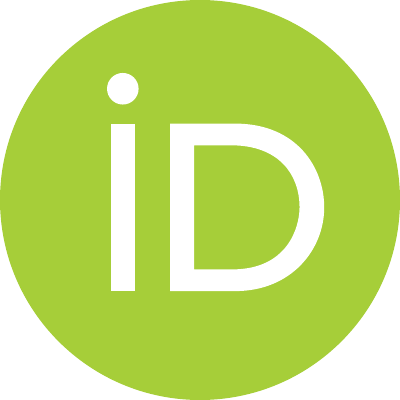}}}}

\let\oldequation\equation
\let\oldendequation\endequation
\renewenvironment{equation}
 {\linenomathNonumbers\oldequation}
 {\oldendequation\endlinenomath}

\begin{document}
\title{\boldmath Observation of the electromagnetic Dalitz transition $J/\psi\to e^+ e^- \eta_c$}
\author{
M.~Ablikim$^{1}$\BESIIIorcid{0000-0002-3935-619X},
M.~N.~Achasov$^{4,c}$\BESIIIorcid{0000-0002-9400-8622},
P.~Adlarson$^{83}$\BESIIIorcid{0000-0001-6280-3851},
X.~C.~Ai$^{89}$\BESIIIorcid{0000-0003-3856-2415},
C.~S.~Akondi$^{31A,31B}$\BESIIIorcid{0000-0001-6303-5217},
R.~Aliberti$^{39}$\BESIIIorcid{0000-0003-3500-4012},
A.~Amoroso$^{82A,82C}$\BESIIIorcid{0000-0002-3095-8610},
Q.~An$^{78,65,\dagger}$,
Y.~H.~An$^{89}$\BESIIIorcid{0009-0008-3419-0849},
M.~S.~Anderson$^{39}$\BESIIIorcid{0009-0008-1550-2632}
% ...
\\
\vspace{0.2cm}
(BESIII Collaboration)\\
\vspace{0.2cm} {\it
$^{1}$ Institute of High Energy Physics, Beijing 100049, People's Republic of China\\
% ...
$^{4}$ Budker Institute of Nuclear Physics SB RAS (BINP), Novosibirsk 630090, Russia\\
% ...
$^{31}$ INFN Sezione di Ferrara, (A)INFN Sezione di Ferrara, I-44122, Ferrara, Italy; (B)University of Ferrara, I-44122, Ferrara, Italy\\
% ...
$^{39}$ Johannes Gutenberg University of Mainz, Johann-Joachim-Becher-Weg 45, D-55099 Mainz, Germany\\
% ...
$^{65}$ State Key Laboratory of Particle Detection and Electronics, Beijing 100049, Hefei 230026, People's Republic of China\\
% ...
$^{78}$ University of Science and Technology of China, Hefei 230026, People's Republic of China\\
% ...
$^{82}$ University of Turin and INFN, (A)University of Turin, I-10125, Turin, Italy; (B)University of Eastern Piedmont, I-15121, Alessandria, Italy; (C)INFN, I-10125, Turin, Italy\\
$^{83}$ Uppsala University, Box 516, SE-75120 Uppsala, Sweden\\
% ...
$^{89}$ Zhengzhou University, Zhengzhou 450001, People's Republic of China\\
\vspace{0.2cm}
$^{\dagger}$ Deceased\\
% ...
$^{c}$ Also at the Novosibirsk State University, Novosibirsk, 630090, Russia\\
% ...
}
}

\begin{abstract}

Using $(10.087\pm0.044)\times10^9$ $J/\psi$ events collected with the BESIII detector at the $e^+e^-$ BEPCII collider, we present the first observation of the electromagnetic Dalitz decay $J/\psi\to e^+ e^- \eta_c$. The relative branching fraction $R \equiv \frac{\Gamma(J/\psi\to e^+e^-\eta_c)}{\Gamma(J/\psi\to \gamma\eta_c)}$ is determined to be $(0.65\pm0.02_{\rm stat.}\pm0.07_{\rm sys.})\%$, where the first uncertainty is statistical and the second systematic. The $q^2$-dependent form factors are also extracted, and no significant deviation from the theoretical prediction is seen.

\end{abstract}

\maketitle

\section{Introduction}

\begin{figure}[htbp]
	\centering
	\includegraphics[width=0.40\textwidth]{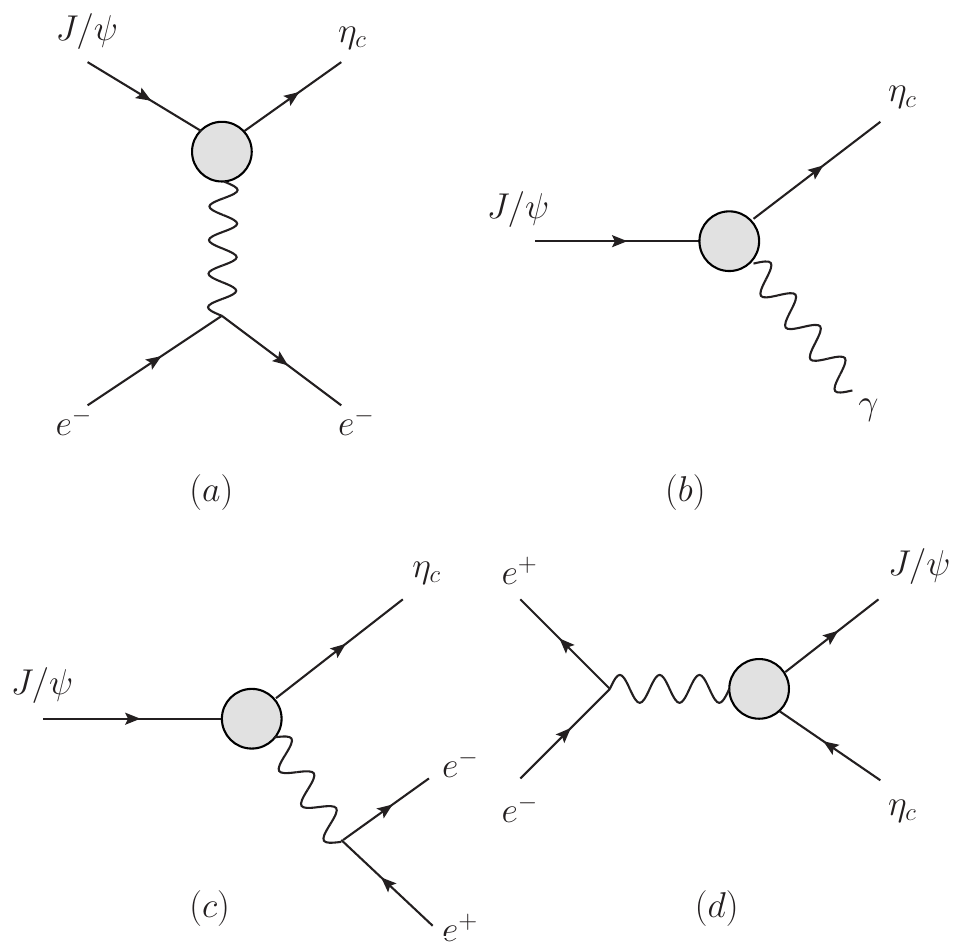}
	\caption{Feynman diagrams of the strong interaction between $J/\psi$ and $\eta_c$ in various processes with different $q^2$ regions: (a) $e^- J/\psi\to e^-\eta_c$ with $q^2<0$; (b) $J/\psi\to\gamma\eta_c$ with $q^2=0$; (c) $J/\psi\to e^+e^-\eta_c$ with $4m_e^2<q^2<(M_{J/\psi}-M_{\eta_c})^2$; (d) $e^+e^-\to J/\psi\eta_c$ with $q^2>(M_{J/\psi}+M_{\eta_c})$.}
	\label{fig:feynman_diag}
\end{figure}

The electromagnetic~(EM) Dalitz decay $J/\psi\to e^+ e^- \eta_c$ provides a unique environment to probe the internal EM structures of the ground-state charmonium states $J/\psi$ and $\eta_c$, and thus contributes to our understanding of the fundamental theory of strong interaction, quantum chromodynamics~(QCD). In general, the interaction between $J/\psi$ and $\eta_c$ can be described by a phenomenological hadronic form factor $F(q^2)$, which can be measured in various processes covering different regions of $q^2=(p_{J/\psi}-p_{\eta_c})^2$, as illustrated in Fig.~\ref{fig:feynman_diag}. Except for the process $e^- J/\psi\to e^-\eta_c$, which is difficult to access experimentally, the reactions $J/\psi\to\gamma\eta_c$~\cite{CLEO:2008pln,Anashin:2010nr,Anashin:2014wva,Beilin:1984pf,Brambilla:2005zw,Pineda:2013lta} and $e^+e^-\to J/\psi\eta_c$~\cite{Belle:2002tfa,Belle:2004abn,BaBar:2005nic,Huang:2022dfw,Sun:2021tma,Zeng:2021hwt} have been extensively studied both experimentally and theoretically. In contrast, there has been no experimental measurement of $J/\psi\to e^+e^-\eta_c$ so far.

Experimentally, several EM Dalitz decays among charmonium states, including $\psi(3686)\to e^+e^-\chi_{cJ}$~\cite{BESIII:2017ung}, $\psi(3686)\to e^+e^-\eta_c$~\cite{BESIII:2022jde}, $\chi_{cJ}\to l^+l^- J/\psi$~\cite{BESIII:2017ung,BESIII:2019yeu}, and $h_c\to e^+e^-\eta_c$~\cite{BESIII:2024kkf}, have been reported by the BESIII collaboration using an unprecedentedly large $\psi(3686)$ data sample. However, the EM Dalitz decay $J/\psi\to e^+e^-\eta_c$ has not yet been observed. To date, the \mbox{BESIII} experiment has collected approximately $10$ billion $J/\psi$ events~\cite{BESIII:2021cxx}, which enables an experimental study of $J/\psi\to e^+e^-\eta_c$ and provides an opportunity to test theoretical predictions based on the Vector-Meson-Dominance Model~(VMD)~\cite{Gell-Mann:1961jim,Bauer:1977iq} and Lattice-QCD calculations~\cite{Colquhoun:2023zbc}.

In this paper, based on $(10.087\pm0.044)\times10^9$ $J/\psi$ events~\cite{BESIII:2021cxx} collected with the BESIII detector, we report the first observation of $J/\psi\to e^+e^-\eta_c$. The relative branching fraction $R \equiv \frac{\Gamma(J/\psi\to e^+e^-\eta_c)}{\Gamma(J/\psi\to \gamma\eta_c)}$ is measured.

\section{DETECTOR AND DATA SAMPLES}
\label{sec:BES}

The BESIII detector~\cite{BESIII:2009fln} records symmetric $e^+e^-$ collisions provided by the BEPCII storage ring~\cite{Yu:2016cof} in the center-of-mass energy ($\sqrt{s}$) range from 2.0 to 4.95~GeV, with a peak luminosity of $1 \times 10^{33}~\text{cm}^{-2}\text{s}^{-1}$ achieved at $\sqrt{s} = 3.77~\text{GeV}$. BESIII has collected large data samples in this energy region~\cite{BESIII:2020nme,Lu:2020imt,Zhang:2022bdc}. The cylindrical core of the BESIII detector covers 93\% of the full solid angle and consists of a helium-based multilayer drift chamber~(MDC), a plastic scintillator time-of-flight system~(TOF), and a CsI(Tl) electromagnetic calorimeter~(EMC), all enclosed within a superconducting solenoidal magnet providing a 1.0~T magnetic field. The magnetic field was 0.9~T in 2012, which affects 10\% of the total $J/\psi$ data. The solenoid is supported by an octagonal flux-return yoke with resistive plate counter muon identification modules interleaved with steel.

The charged-particle momentum resolution at $1~{\rm GeV}/c$ is $0.5\%$, and the ionization energy loss ${\rm d}E/{\rm d}x$ resolution is $6\%$ for electrons from Bhabha scattering. The EMC measures photon energies with a resolution of $2.5\%$ ($5\%$) at $1$~GeV in the barrel (end cap) region. The time resolution in the TOF barrel region is 68~ps, while that in the end cap region is 110~ps. The end cap TOF system was upgraded in 2015 using multigap resistive plate chamber technology, providing a time resolution of 60~ps~\cite{Li:2017jpg,Guo:2017sjt,Cao:2020ibk}, which benefits 87\% of the data used in this analysis.

Monte Carlo (MC) simulated samples are produced with a {\sc geant4}-based~\cite{GEANT4:2002zbu} software package, which includes the geometric description of the BESIII detector and its response. These samples are used to determine detection efficiencies and to estimate backgrounds. The simulation models the beam energy spread and initial-state radiation in $e^+e^-$ annihilations using the generator {\sc kkmc}~\cite{Jadach:1999vf,Jadach:2000ir}. The inclusive MC sample includes both the production of the $J/\psi$ resonance and continuum processes incorporated in {\sc kkmc}~\cite{Jadach:1999vf,Jadach:2000ir}. All particle decays are modelled with {\sc evtgen}~\cite{Lange:2001uf,Ping:2008zz}, using branching fractions taken from the particle data group~(PDG)~\cite{ParticleDataGroup:2024cfk}, when available, or otherwise estimated with {\sc lundcharm}~\cite{Chen:2000tv,Yang:2014vra}. Final-state radiation from charged particles is simulated using the {\sc photos} package~\cite{Barberio:1990ms}.

The exclusive MC sample of $J/\psi\to e^+e^-\eta_c$ is generated based on the amplitude from Refs.~\cite{Gu:2019qwo,BESIII:2018aao}. The non-negligible width of $\eta_c$~\cite{ParticleDataGroup:2024cfk}, as well as the corresponding damping factor in the KEDR form~\cite{Anashin:2010nr}, are incorporated into this amplitude. The same damping factor is also applied to the exclusive MC sample of $J/\psi\to \gamma\eta_c$. All $\eta_c$ hadronic decays are simulated using a phase-space generator.

\section{EVENT SELECTION}
\label{sec:selection}

In this work, the radiative photon from $J/\psi\to\gamma\eta_c$ is reconstructed via the $\gamma$-conversion process. Since both the EM Dalitz decay $J/\psi\to e^+e^-\eta_c$ and $J/\psi\to\gamma\eta_c$ lead to exactly the same final states, common sources of systematic uncertainties cancel in the determination of the relative branching fraction $R$. The $\eta_c$ is reconstructed through 14 hadronic decay modes: $K_S^0 K^\pm\pi^\mp$, $K^+K^-\pi^0$, $K^+K^-\eta$, $\pi^+\pi^-\eta$, $\pi^+\pi^-\pi^+\pi^-\eta$, $K^+K^-\pi^+\pi^-\pi^0$, $K_S^0K^\pm\pi^\mp\pi^\pm\pi^\mp$, $\pi^+\pi^-\pi^+\pi^-\pi^+\pi^-$, $K^+K^-\pi^+\pi^-$, $K^+K^-\pi^+\pi^-\pi^+\pi^-$, $\pi^+\pi^-\pi^+\pi^-$, $p\bar{p}$, $p\bar{p}\pi^0$, and $p\bar{p}\pi^+\pi^-$. These modes cover approximately 22\% of the total $\eta_c$ decay width. Although the decay modes $\pi^+\pi^-\pi^0\pi^0$ and $\pi^+\pi^-\pi^+\pi^-\pi^0\pi^0$ have relatively large branching fractions, they are not considered in this analysis due to the overwhelmingly large background contributions from $J/\psi$ decays into multiple pions.
Charged tracks are required to originate from the region defined by $|V_z|<10$ cm, $V_r<1$ cm, and $|\cos\theta|<0.93$, where $|V_z|$ and $V_r$ denote the distances of closest approach of the reconstructed track to the interaction point~(IP) along the $z$ direction and in the transverse $(x,y)$ plane, respectively. Here, $\theta$ is the polar angle of the charged track with respect to the $z$-axis, which is defined as the symmetry axis of the MDC.

To identify $\pi^\pm,K^\pm,p^\pm$ from the charged tracks, particle identification~(PID) is performed by using the TOF and d$E$/dx information. The PID confidence level~($CL$) is calculated for all three hypotheses, and we require
\begin{itemize}
	\item $CL_{\pi}>CL_{K}$ and $CL_\pi>0$ for the $\pi^\pm$ tracks.
	\item $CL_{K}>CL_{\pi}$, $CL_{K}>CL_{p}$, and $CL_{K}>0$ for the $K^\pm$ tracks.
	\item $CL_{p}>CL_{\pi}$ and $CL_{p}>CL_{K}$ for the $p^\pm$ tracks.
\end{itemize}
No PID requirement is imposed on the $e^\pm$ tracks due to their relatively low momenta.


The $K_S^0$ candidates are reconstructed via $K_S^0\to \pi^+\pi^-$. Due to the relatively long lifetime of the $K_S^0$, the charged pions from its decay do not originate from the IP. Therefore, the requirement on $|V_z|$ is relaxed to be less than 20 cm, and no requirement is imposed on $V_r$, while the polar angle requirement $|\cos\theta|<0.93$ is retained. All pairs of oppositely charged tracks that satisfy the above criteria are assigned as $\pi^+\pi^-$ candidates and are constrained to originate from a common vertex using a vertex fit. The decay length of each $K_S^0$ candidate, defined as the distance between the IP and the decay vertex, is required to be greater than twice the vertex resolution. The invariant mass of the $\pi^+\pi^-$ pair is required to satisfy $|M_{\pi^+\pi^-}-M_{K_S^0}|<12$ MeV/$c^2$, where $M_{K_S^0}$ is the nominal mass of the $K_S^0$~\cite{ParticleDataGroup:2024cfk}.

The $\pi^0$ and $\eta$ candidates are reconstructed via their two-photon decays, where photon candidates are identified as isolated showers in the EMC. The deposited energy of each shower is required to be at least 25 MeV in the barrel region and 50 MeV in the end-cap regions. To suppress final-state radiation from charged tracks, the angle between the shower position and any extrapolated charged track in the EMC must be greater than $10^\circ$. The difference between the EMC time and the event start time is required to be within [0,~700]~ns to suppress electronic noise and energy deposits from other events. For the selected photon pairs, invariant mass windows of $M_{\gamma\gamma}\in(0.115,0.150)$ GeV/$c^2$ and $M_{\gamma\gamma}\in(0.500,0.570)$ GeV/$c^2$ are required for $\pi^0$ and $\eta$ candidates, respectively. In addition, a one-constraint~(1C) kinematic fit is performed by constraining the invariant mass of the two photons to the known $\pi^0$ or $\eta$ mass, with the requirement $\chi^2_{\rm 1C}<50$.

For each hadronic decay mode $\eta_c\to f$, a four-constraint~(4C) kinematic fit is performed for the selected final states $e^+e^- f$ by imposing overall energy–momentum conservation. The combination with the smallest $\chi^2_{\rm 4C}$ is retained for further analysis. The requirement on $\chi^2_{\rm 4C}$ is optimized to be less than 15. In addition, the invariant mass of the $e^+e^-$ pair is required to be less than 0.5 GeV/$c^2$, and events with extra charged tracks are rejected.

\begin{figure}[htbp]
	\centering
	\includegraphics[width=8cm]{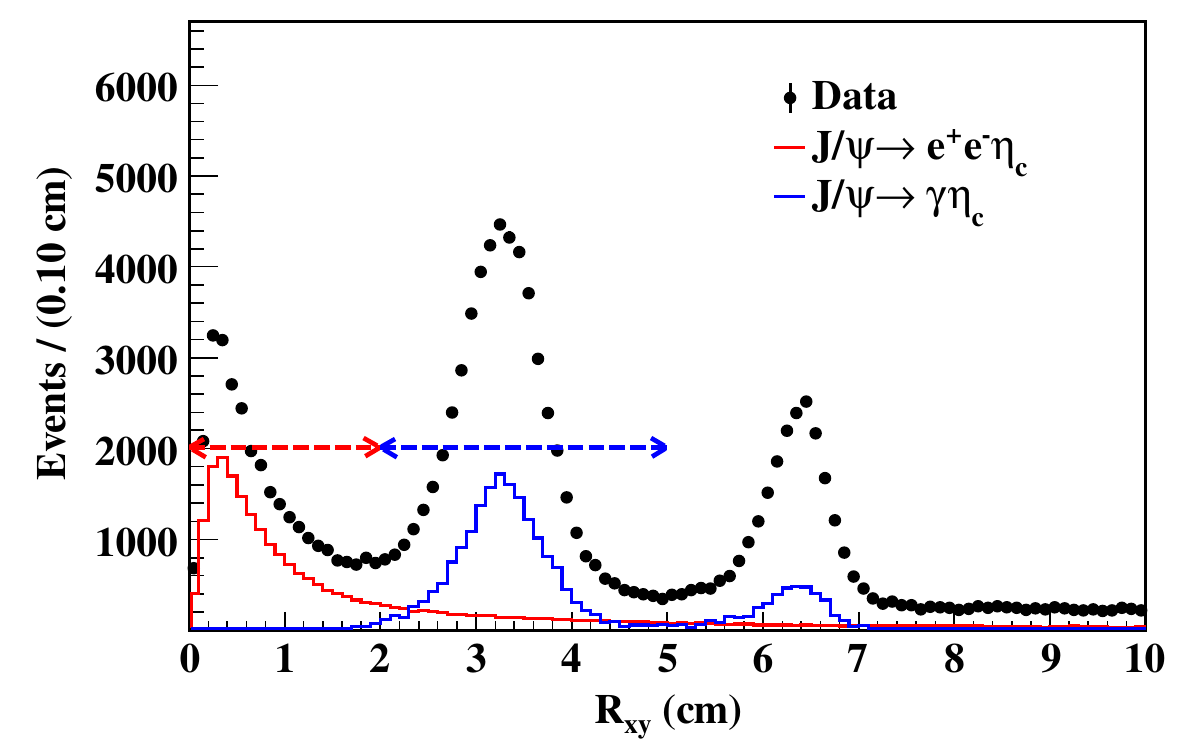}
	\caption{Distributions of $R_{xy}$ of the $e^+e^-$ pairs. The three peaks correspond to pairs originating from the interaction point, the beam pipe, and the MDC inner wall. The dots with error bars are the data sample. The red and blue solid histograms are the MC samples of $J/\psi\to e^+ e^-\eta_c$ and $J/\psi\to \gamma\eta_c$ with arbitrary normalization, respectively. The dashed red and blue arrows indicate the signal region I and II.}
	\label{fig:Rxy}
\end{figure}

To distinguish $e^+e^-$ pairs originating from $\gamma$-conversion and EM Dalitz decay, an algorithm is applied to identify $\gamma$-conversion. The $\gamma$-conversion point~(CP) is reconstructed from the intersection of the two charged-track trajectories in the $(x,y)$ plane. The $\gamma$-conversion length $R_{xy}$ is defined as the distance from the beam axis to the CP in the $(x,y)$ plane. As shown in Fig.~\ref{fig:Rxy}, three peaks are observed in the regions $0<R_{xy}<2$~cm, $2<R_{xy}<5$~cm, and $5<R_{xy}<8$~cm, corresponding to $e^+e^-$ pairs originating from the interaction point, $\gamma$-conversion in the beam pipe, and $\gamma$-conversion in the MDC inner wall, respectively. The processes of interest, $J/\psi\to e^+ e^-\eta_c$ and $J/\psi\to \gamma\eta_c$, are primarily located in region I ($0<R_{xy}\leq2$~cm) and region II ($2<R_{xy}<5$~cm), respectively. Since only a limited number of signal events are observed for $R_{xy}>5$~cm, a requirement of $R_{xy}<5$~cm is imposed in the analysis.

\section{BACKGROUND STUDY}

\begin{figure*}[htbp]
	\centering
	\centering
	\setlength{\abovecaptionskip}{-0.0cm}
	\includegraphics[width=17.0cm]{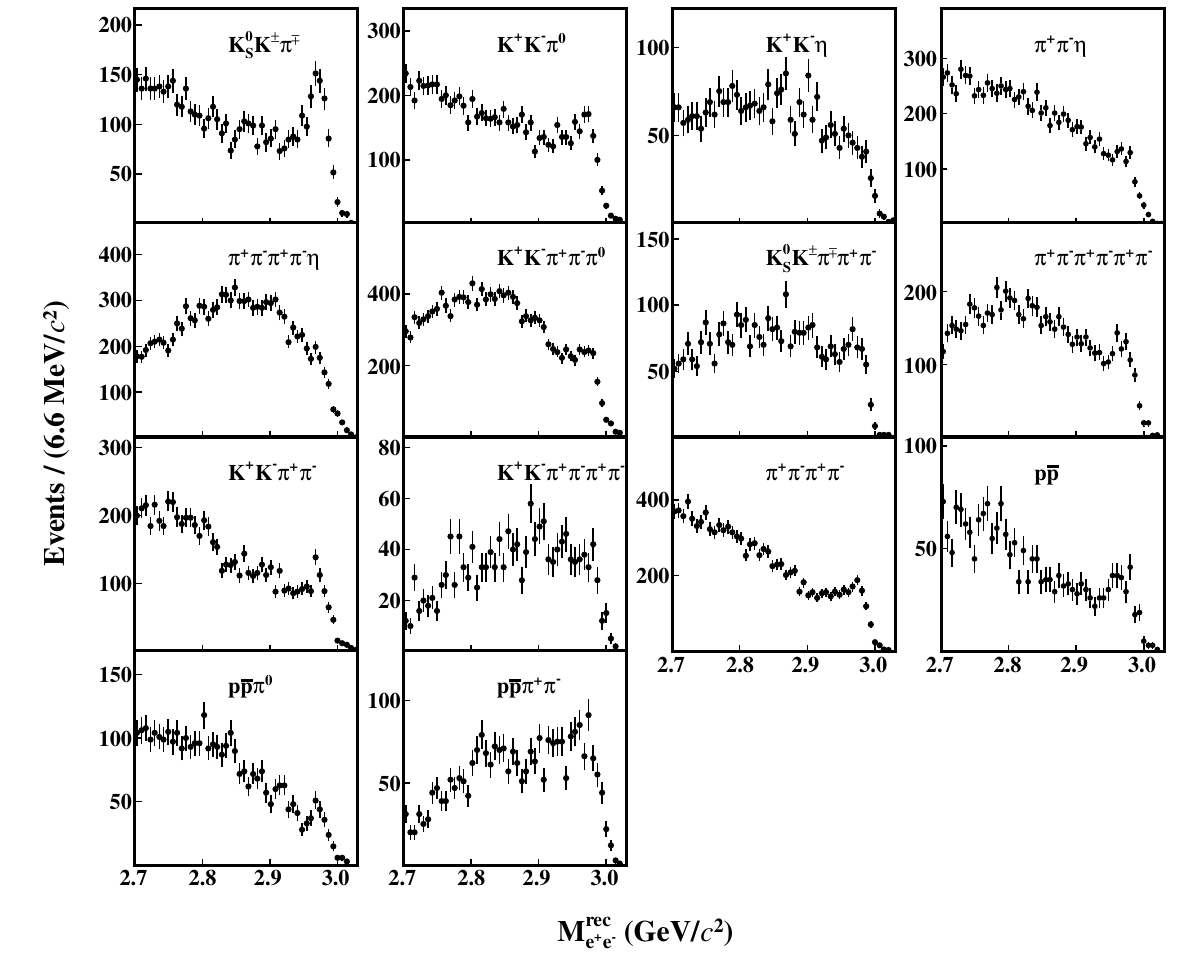}
	\caption{Distributions of the $e^+e^-$ recoil mass, $M^{\rm rec}_{e^+e^-}$, for all 14 hadronic decay modes of $\eta_c$ in data.}
	\label{fig:mass_rec_ee_eachmode}
\end{figure*}

After applying all the above selection criteria, Fig.~\ref{fig:mass_rec_ee_eachmode} shows the distributions of the invariant mass recoiling against the $e^+e^-$ pair, $M^{\rm rec}_{e^+e^-}$, for all 14 hadronic decay modes of the $\eta_c$. Significant $\eta_c$ signal contributions are observed in most of the decay modes.

In principle, three types of potential $\eta_c$ peaking backgrounds can contribute:
\begin{itemize}
\item Other $\eta_c$ production in $J/\psi$ decays. Given the limited available phase space, no significant processes are expected other than $J/\psi\to\gamma\eta_c$ and $J/\psi\to e^+e^-\eta_c$.
\item $\eta_c$ decays into background hadronic modes $f_{\rm bkg}$ that are reconstructed as signal modes $f_{\rm sig}$. For example, for the signal decay $\eta_c \to K_S^0 K^\pm \pi^\mp$, the background process $\eta_c \to K_S^0 K^\pm \pi^\mp \pi^0$ may be reconstructed as the signal channel if a soft $\pi^0$ is not detected, thereby producing a peak in the $M^{\rm rec}_{e^+e^-}$ distribution. However, since this type of background contributes to both $J/\psi\to\gamma\eta_c$ and $J/\psi\to e^+e^-\eta_c$, it can be treated as part of the signal and does not affect the measurement of the relative branching fraction.
\item $\eta_c$ production in continuum processes. In quantum electrodynamics~(QED), $\eta_c$ production is dominated by two-photon processes. Given the small branching fraction $\mathcal{B}(\eta_c\to\gamma\gamma)=(1.66\pm0.13)\times10^{-4}$~\cite{ParticleDataGroup:2024cfk}, the corresponding contribution is negligible. This is further validated using an off-resonance data sample at $\sqrt{s}=3.080$~GeV with an integrated luminosity of $(167.06\pm0.10)$~pb$^{-1}$~\cite{BESIII:2021cxx,BESIII:2016kpv}, where no significant $\eta_c$ signal is observed.
\end{itemize}

Therefore, potential $\eta_c$ peaking backgrounds are neglected in the subsequent signal extraction.

The non-peaking background contributions are mainly dominated by misidentification of final states, missing soft photons from $\pi^0$ decays, as well as photons from final-state radiation and non-$\eta_c$ radiative decays of the $J/\psi$.

\section{RELATIVE BRANCHING FRACTION}

\subsection{Signal extraction}

\begin{figure}[htbp]
	\centering
	\includegraphics[width=7cm]{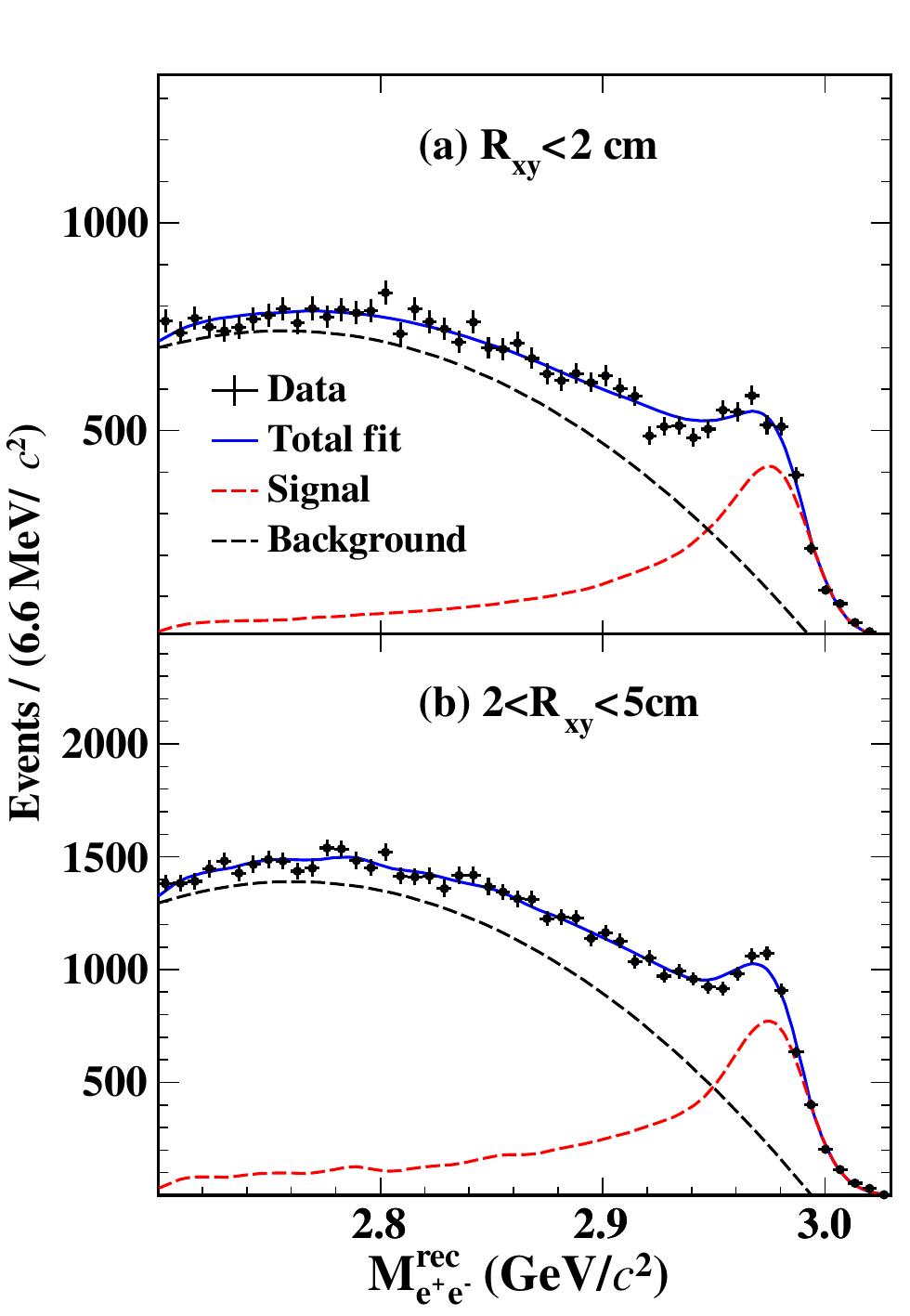}
	\caption{Fit results to the $M^{\rm rec}_{e^+e^-}$ distributions in the signal regions (a)~I and (b)~II, where the contributions from all 14 decay modes are overlapped. The dots with error bars are from data sample. The solid blue lines are the total fit results. The black and red dashed lines are the signal and background contributions, respectively.}
	\label{fig:fit_to_Mrec}
\end{figure}

The signal yields of $\eta_c$ are extracted by performing fits to the combined $M^{\rm rec}_{e^+e^-}$ distributions in the range $[2.70,3.03]$~GeV/$c^2$ for the signal regions I and II separately. The signal shape is described by the MC-simulated distribution convolved with a Gaussian function with free parameters to account for possible resolution differences between data and simulation. The background is modeled using a second-order polynomial function. The fit results are shown in Fig.~\ref{fig:fit_to_Mrec}. The extracted signal yields are $N^{\rm obs}_{\rm I}=6071\pm158$ and $N^{\rm obs}_{\rm II}=11320\pm235$ for regions I and II, respectively, where the uncertainties are statistical only.

The potential interference between the $\eta_c$ signal and non-resonant components is neglected in the fit due to the overwhelmingly large background level. Since the form factor $F(q^2)$ is generally considered to be real~\cite{Colquhoun:2023zbc}, no additional phase is expected between the processes $J/\psi\to\gamma \eta_c$ ($q^2=0$) and $J/\psi\to e^+e^-\eta_c$ ($q^2>0$). Therefore, any uncertainty arising from the neglected interference is expected to largely cancel in the measurement of the relative branching fraction.

\subsection{Efficiency matrix}

Since the requirement on $R_{xy}$ cannot fully distinguish between the EM Dalitz decay and $\gamma$-conversion, both processes contribute to the signal regions I and II, as shown in Fig.~\ref{fig:Rxy}. Based on MC simulations, the corresponding efficiency matrix is defined as
\begin{equation}
\begin{bmatrix}
\epsilon_{{ee}\to {\rm I}} &  \epsilon_{{\gamma}\to {\rm I}}\\
\epsilon_{{ee}\to {\rm II}} & \epsilon_{{\gamma}\to {\rm II}}
\end{bmatrix}.
\end{equation}
Here, the matrix element $\epsilon_{a\to b}$ represents the reconstruction efficiency for process $a=ee,\gamma$ with the corresponding $R_{xy}$ value falling within signal region $b={\rm I,II}$.

\subsection{Correction on efficiency matrix}


The efficiency matrix strongly depends on the MC-simulated $R_{xy}$ distributions for both the EM Dalitz decay and the $\gamma$-conversion process. In this work, the EM Dalitz decay $\pi^0\to \gamma e^+e^-$ and the process $\pi^0\to\gamma\gamma$ with a single $\gamma$ converting into an $e^+e^-$ pair are used as control samples to evaluate potential deviations. In these samples, the $\pi^0$ originates from the process $J/\psi\to\pi^+\pi^-\pi^0$.

For the control sample, the charged tracks ($e^\pm,\pi^\pm$) and neutral showers ($\gamma$) are selected using the same criteria as described in Sec.~\ref{sec:selection}. A similar four-constraint (4C) kinematic fit is applied, with the requirement $\chi^2_{\rm 4C}<50$. To ensure that the $e^+e^-$ pairs populate the same momentum region as those from $J/\psi\to e^+e^-\eta_c$, an additional requirement of $M^{\rm rec}_{e^+e^-}>2.7$~GeV/$c^2$ is imposed. To suppress non-$\pi^0$ backgrounds, the invariant mass is required to satisfy $|M_{e^+e^-\gamma}-M_{\pi^0}|<0.03$~GeV/$c^2$, where $M_{\pi^0}$ is the nominal $\pi^0$ mass~\cite{ParticleDataGroup:2024cfk}.

\begin{figure}[htbp]
	\centering
	\includegraphics[width=0.47\textwidth]{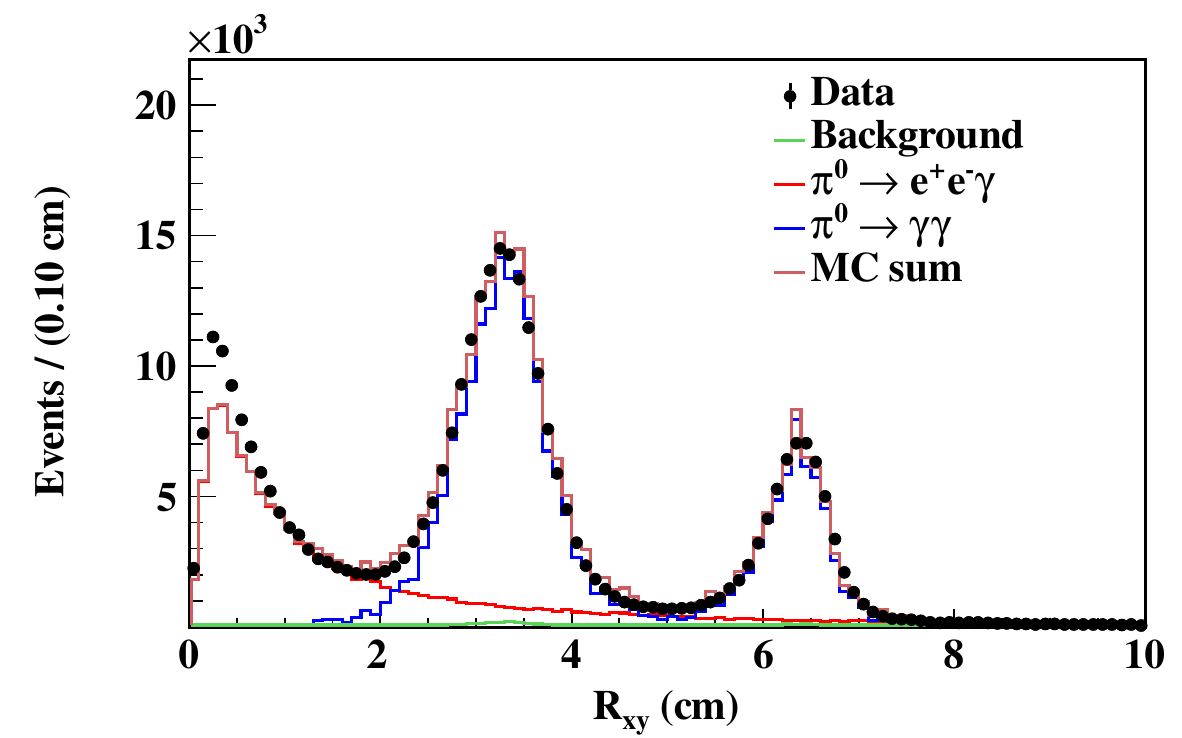}
	\caption{Distributions of $R_{xy}$ of the $e^+e^-$ pairs of the accepted candidates for the control sample after global scaling and before $R_{xy}$ correction. The dots with error bars are the data sample. The red and blue solid histograms are the MC samples of $\pi^0\to e^+ e^-\gamma$ and $\pi^0\to\gamma\gamma$ with normalization described in main text, respectively. The gray solid histogram is the simulated background.}
	\label{fig:Rxy_controlsample}
\end{figure}

Figure~\ref{fig:Rxy_controlsample} shows the $R_{xy}$ distribution for the control sample. The MC samples of $J/\psi\to\pi^+\pi^-\pi^0$ are generated based on amplitude analysis results and normalized using branching fractions from the PDG~\cite{ParticleDataGroup:2024cfk}. An additional global scaling is applied to the MC distributions to account for residual differences in tracking and PID efficiencies. This scaling does not introduce any bias in the measurement of the relative branching fraction. As shown in Fig.~\ref{fig:Rxy_controlsample}, the MC modeling of the $\gamma$-conversion process is in good agreement with the data; therefore, no correction is applied to $\epsilon_{{\gamma}\to {\rm I}}$ and $\epsilon_{{\gamma}\to {\rm II}}$.

For the EM Dalitz decay, however, some discrepancies are observed. Correction factors are therefore introduced as
\begin{equation}
\begin{split}
f_{{ee}\to {\rm I}} &= \frac{N_{\rm data}(R_{xy}<2)-N_{\rm bg}(R_{xy}<2)}{N_{\pi^0\to \gamma e^+ e^-}(R_{xy}<2)},\\
f_{{ee}\to {\rm II}} &=  \frac{N_{\rm data}(2<R_{xy}<5)-N_{\rm bg}(2<R_{xy}<5)}{N_{\pi^0\to \gamma e^+ e^-}(2<R_{xy}<5)}, \\
\end{split}
\end{equation}
which are determined to be $1.12\pm0.01$ and $0.64\pm0.10$, respectively. The efficiency matrix elements $\epsilon_{{ee}\to {\rm I}}$ and $\epsilon_{{ee}\to {\rm II}}$ are then corrected to $\epsilon^\prime_{{ee}\to {\rm I}}=f_{{ee}\to {\rm I}}\times\epsilon_{{ee}\to {\rm I}}$ and $\epsilon^\prime_{{ee}\to {\rm II}}=f_{{ee}\to {\rm II}}\times\epsilon_{{ee}\to {\rm II}}$, respectively.

\subsection{Relative branching fraction}

The corrected efficiency matrix elements are given as $\epsilon^\prime_{{ee}\to {\rm I}}=(2.10\pm0.01)\%$, $\epsilon^\prime_{{\gamma}\to {\rm I}}=(4.4\pm0.4)\times10^{-6}$, $\epsilon^\prime_{{ee}\to {\rm II}}=(0.31\pm0.01)\%$, and $\epsilon^\prime_{{\gamma}\to {\rm II}}=(2.44\pm0.03)\times10^{-4}$. The number of produced events are determined with
\begin{equation}
\begin{bmatrix}
{ N^{\rm pro}_{ee}}\\
{N^{\rm pro}_{\gamma}}
\end{bmatrix}
=
\begin{bmatrix}
\epsilon^\prime_{{ee}\to {\rm I}} &  \epsilon^\prime_{{\gamma}\to {\rm I}}\\
\epsilon^\prime_{{ee}\to {\rm II}} & \epsilon^\prime_{{\gamma}\to {\rm II}} 
\end{bmatrix}^{-1}
\cdot
\begin{bmatrix}
{ N^{\rm obs}_{\rm I}}\\
{N^{\rm obs}_{\rm II}}
\end{bmatrix}.
\end{equation}
Finally, one obtains the relative branching fraction as
\begin{equation}
R \equiv \frac{\Gamma(J/\psi\to e^+e^-\eta_c)}{\Gamma(J/\psi\to \gamma\eta_c)} = \frac{N^{\rm pro}_{ee}}{N^{\rm pro}_{\gamma}},
\end{equation}
which is determined to be $(0.65\pm0.02)\%$, where the uncertainty is statistical only. The statistical uncertainty is evaluated by varying $N^{\rm obs}_{\rm I}$ and $N^{\rm obs}_{\rm II}$ according to their respective statistical uncertainties, repeating the calculation of $R$ 1000 times, and taking the standard deviation of the resulting distribution as the statistical uncertainty. Further systematic uncertainties related to efficiency matrix are also estimated in the similar approach by generating alternative efficiency matrices.

\section{SYSTEMATIC UNCERTAINTIES}

Since the EM Dalitz decay and the $\gamma$-conversion process lead to exactly the same final states, most systematic uncertainties cancel, including those related to tracking and PID of charged tracks, reconstruction of $\pi^0$, $\eta$, and $K^0_S$, and the MC modeling of $\eta_c$ hadronic decays. The remaining sources of systematic uncertainty are summarized in Table~\ref{tab:sys_table}. The total systematic uncertainty is obtained by adding the individual contributions in quadrature.

\begin{table}[htbp]
	\centering
	\tabcolsep=0.6cm
	\caption{The systematic uncertainties of the relative branching fraction measurement.}
	\begin{tabular}{cc}
		\hline\hline
		Source & Uncertainty~(\%)\\
		\hline
		MC statistics &  1.8\\
		Correction factor & 6.5\\
		Fit to $M^{\rm rec}_{e^+e^-}$ & 7.6\\
		MC modeling & --- \\
		Insignificant channel & 0.9\\
		Kinematic fit & 1.2\\
		\hline
		Sum & 10.3\\
		\hline
		\hline
	\end{tabular}
	\label{tab:sys_table}
\end{table}

\subsection{MC statistics}
Due to the relatively low detection efficiency, the systematic uncertainty arising from MC statistics, given by $\sqrt{\frac{1-\epsilon}{N\cdot\epsilon}}$, is non-negligible even with a large MC sample. Here, $\epsilon$ denotes the detection efficiency and $N$ the number of generated events. The corresponding impact on the relative branching fraction is estimated by generating 1000 alternative efficiency matrices, where the matrix elements are varied according to their statistical uncertainties. The relative branching fraction is recalculated for each case, and the standard deviation of the resulting distribution is taken as the systematic uncertainty.

\subsection{Correction factor}
The systematic uncertainty associated with the correction factors applied to the efficiency matrix elements is estimated by varying these factors. Since the origin of the observed deviations is not fully understood, a conservative approach is adopted: the correction factors are varied by half of their deviations from unity, namely $(1.12-1)/2=0.06$ for 1.12 and $(1-0.64)/2=0.18$ for 0.64. Based on these variations, 1000 alternative efficiency matrices are generated, and the relative branching fraction is recalculated. The standard deviation of the resulting distribution is assigned as the corresponding systematic uncertainty.

\subsection{Fit to $M^{\rm rec}_{e^+e^-}$}
The systematic uncertainty associated with the fit to the $M^{\rm rec}{e^+e^-}$ distribution includes three contributions:
\begin{itemize}
\item \textit{Fit range}: The uncertainty is estimated by varying the upper bound of the fit range to 3.02 GeV/$c^2$ or 3.04 GeV/$c^2$, and the lower bound to 2.68 GeV/$c^2$ or 2.72 GeV/$c^2$.
\item \textit{Signal shape}: The uncertainty is evaluated using an alternative damping factor of the CLEO-c form, $f_d=\exp{(-\frac{E^2\gamma}{8\beta^2})}$ with $\beta=65.0$ MeV~\cite{CLEO:2008pln}.
\item \textit{Background shape}: The uncertainty is estimated by replacing the second-order polynomial with a third-order polynomial function.
\end{itemize}
For each variation, the maximum deviation from the nominal result is taken as the corresponding uncertainty, and the total uncertainty is obtained by adding these contributions in quadrature.

\subsection{MC modeling}
The systematic uncertainty due to the MC modeling of $J/\psi\to e^+e^-\eta_c$ is evaluated by varying the pole mass $\Lambda$ in the form factor $F(q^2)=1/(1-q^2/\Lambda^2)$~\cite{Gu:2019qwo,BESIII:2018aao}. The pole mass is varied from $\Lambda=3.686$ GeV to 3.773 GeV and to infinity. The resulting impact is found to be negligible due to the limited $q^2$ range accessible in this analysis.

\subsection{Insignificant channels}
The systematic uncertainty associated with insignificant $\eta_c$ hadronic decay channels, including $\pi^+\pi^-\pi^+\pi^-\eta$, $K^+K^-\eta$, $\pi^+\pi^-\eta$, and $K^+K^-\pi^+\pi^-\pi^+\pi^-$, is estimated by excluding these channels from the fit. The difference between the nominal and alternative results is taken as the corresponding uncertainty.

\subsection{Kinematic fit}
Since the $\gamma$-conversion process does not strictly conserve the four-momentum of the $e^+e^-$ pair, the systematic uncertainty associated with the 4C kinematic fit does not fully cancel. This effect is evaluated by applying helix parameter corrections~\cite{BESIII:2012mpj} to the $e^\pm$ tracks. An alternative efficiency matrix is obtained, and the difference between the nominal and corrected results is taken as the systematic uncertainty.

\section{Transition Form factors}

To extract the $q^2$-dependent form factors, the data sample is divided into four $M_{e^+e^-}$ intervals, as listed in Table~\ref{tab:FF_measurement}. Compared to the nominal selection criteria, an additional requirement of $R_{xy}<2$ cm is imposed to suppress the $\gamma$-conversion background. Furthermore, to ensure sufficient signal purity in each interval, only the hadronic decay modes $K_S^0 K^\pm\pi^\mp$, $K^+K^-\pi^0$, $\pi^+\pi^-\pi^+\pi^-\pi^+\pi^-$, $K^+K^-\pi^+\pi^-$, $\pi^+\pi^-\pi^+\pi^-$, and $p\bar{p}$ are used in this partial decay rate measurement.

\begin{figure}[htbp]
	\centering
	\vspace{-0.3cm}
	\includegraphics[width=0.47\textwidth]{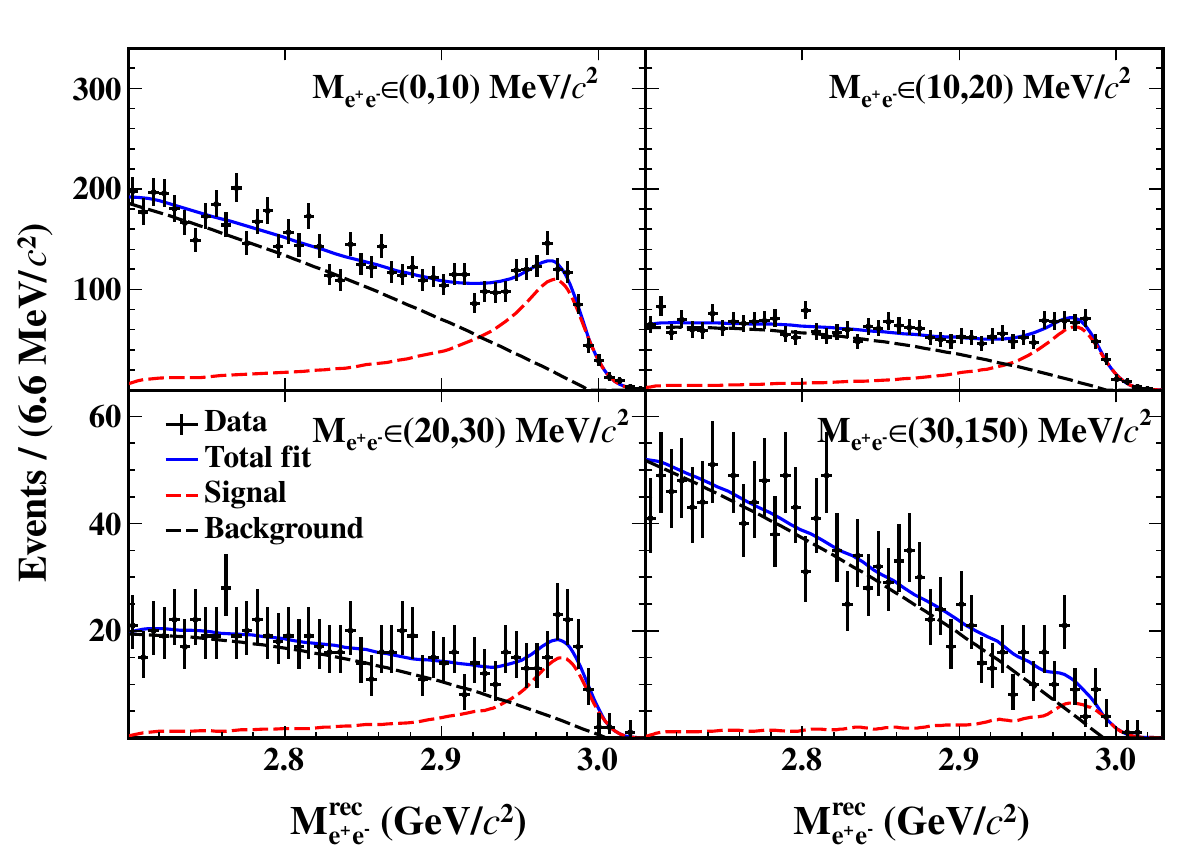}
	\caption{Fit results to $M^{\rm rec}_{e^+e^-}$ in different $M_{e^+e^-}$ intervals. The dots with error bars are from data sample. The solid blue lines are the total fit results. The black and red dashed lines are the signal and background contributions, respectively.}
	\label{fig:Mrec_ee_fit}
\end{figure}

Figure~\ref{fig:Mrec_ee_fit} shows the fit results to $M^{\rm rec}_{e^+e^-}$ in the different $M_{e^+e^-}$ intervals, and the corresponding fitted signal yields $N_{\rm fit}$ are summarized in Table~\ref{tab:FF_measurement}. Although the requirement $R_{xy}<2$ cm is applied, a non-negligible $\gamma$-conversion background remains. Its contribution is subtracted from the fitted yields using $N_{\rm obs}=N_{\rm fit}\times (1-r_{\gamma-{\rm conv}})$. The ratio $r_{\gamma-{\rm conv}}$ is defined as
\begin{equation}
{N^{\rm MC}_{\gamma-{\rm conv}}}/({N^{\rm MC}_{\gamma-{\rm conv}}+f_{ee \to {\rm I}}\times N^{\rm MC}_{{e^+e^-}}}),
\end{equation}
where $N^{\rm MC}_{\gamma-{\rm conv}}$ and $N^{\rm MC}_{e^+e^-}$ denote the numbers of selected $\gamma$-conversion and EM Dalitz MC events, respectively, in the region $R_{xy}<2$ cm.

To account for migration of events between the $M_{e^+e^-}$ intervals due to resolution and mis-reconstruction, the numbers of produced events are calculated as $N^{i}_{\rm pro} = \sum_{j}\epsilon_{ij}^{-1}N^{j}_{\rm obs}$. Here, the efficiency matrix element is defined as $\epsilon_{ij}=N^{\rm rec}_{ij}/N^{\rm gen}_{j}$, where $N^{\rm rec}_{ij}$ is the number of events generated in the $j$-th interval and reconstructed in the $i$-th interval, and $N^{\rm gen}_{j}$ is the number of generated events in the $j$-th interval.

\begin{table}[htbp]
	\centering
	\caption{The $M_{e^+e^-}$ bin division, fitted signal yields $N_{\rm fit}$, and the normalized form factors $F^i_{\rm norm}$ in each interval. Here, the first uncertainties are statistical and the second are systematic.}
	\scalebox{0.9}{
	\begin{tabular}{cccc}
		\hline\hline
		$M_{e^+e^-}$ (MeV/$c^2$) & $N_{\rm fit}$ & $r_{\gamma-{\rm conv}}$~(\%) & $F^i_{\rm norm}$\\
		\hline
		$(0,10)$ & $1701\pm82$ & $0.8\pm0.3$ & 1.0\\
		$(10,20)$ & $804\pm52$ & $4.9\pm1.1$ & $1.16\pm0.16\pm0.39$\\
		$(20,30)$ & $191\pm25$ & $5.0\pm2.5$ & $1.34\pm0.45\pm0.33$\\
		$(30,150)$ & $111\pm24$ & $13.2\pm5.0$ & $1.02\pm0.45\pm0.34$\\
		\hline
		\hline
	\end{tabular}
}
	\label{tab:FF_measurement}
\end{table}

\begin{figure}[htbp]
	\centering
	\vspace{0.0cm}
	\includegraphics[width=0.47\textwidth]{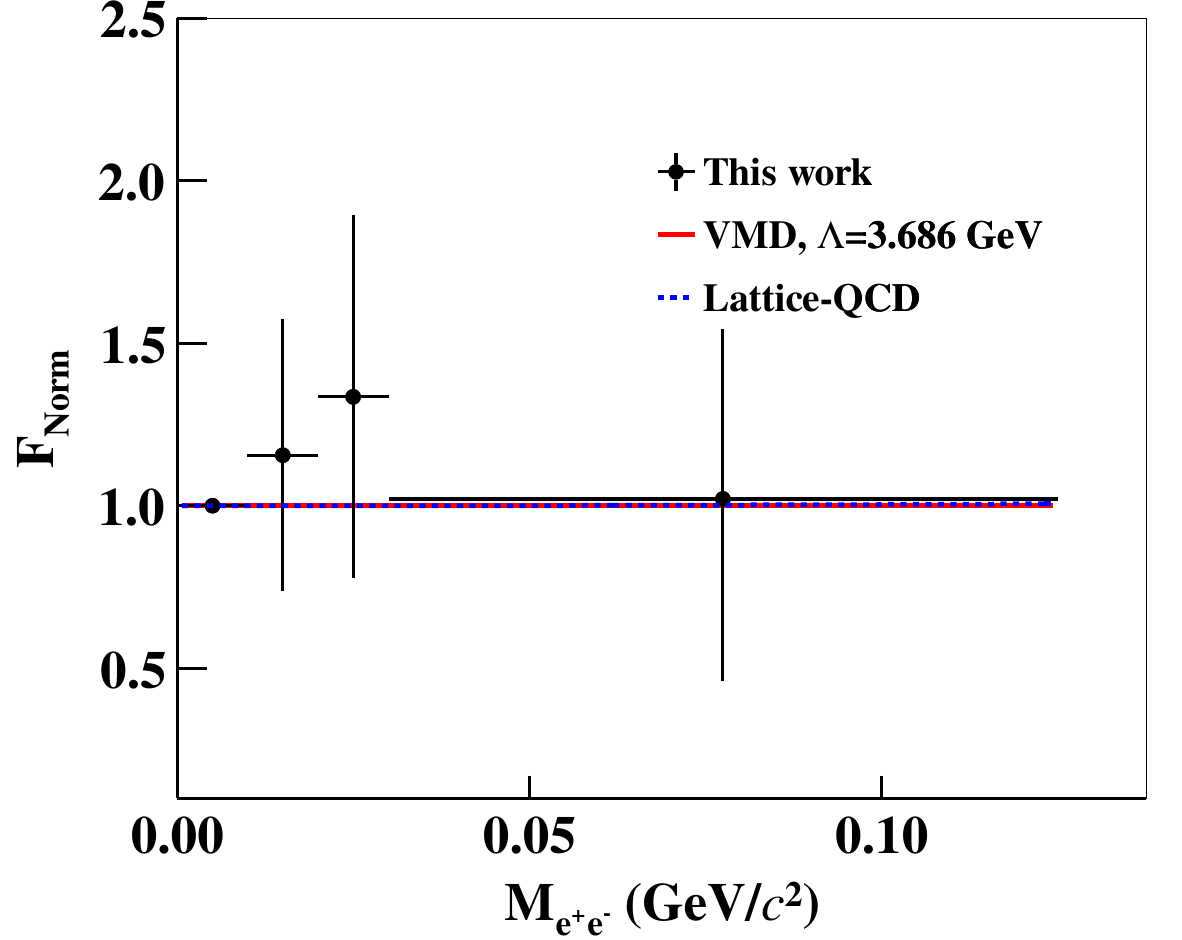}
	\caption{Normalized form factor $F^i_{\rm norm}$ as a function of $M_{e^+e^-}$. The dots with error bars are our measurement results. The red solid line is the VMD predicted lineshape $1/(1-q^2/\Lambda^2)$ with $\Lambda=3.686$ GeV. The blue dashed line is the Lattice QCD calculation result~\cite{Colquhoun:2023zbc}.}
	\label{fig:FFs}
\end{figure}

The $M_{e^+e^-}$ binned form factor $F^i$ is extracted with
\begin{equation}
\begin{split}
F^i &= \sqrt{\frac{\int{\frac{d\Gamma}{d M_{e^+e^-}}\times F^2(M_{e^+e^-})\times d M_{e^+e^-}}}{\int{\frac{d\Gamma}{d M_{e^+e^-}}\times d M_{e^+e^-}}}} \\
&= \sqrt{\frac{c\times N^{i}_{\rm pro}}{\int{\frac{d\Gamma}{d M_{e^+e^-}}\times d M_{e^+e^-}}}}.
\end{split}
\end{equation}
Here, $c$ is an overall constant and $\frac{d\Gamma}{d M_{e^+e^-}}$ is the pure QED contribution without form factor~\cite{Gu:2019qwo},
\begin{equation}
\begin{split}
\frac{d\Gamma}{d M_{e^+e^-}}&=2\Gamma(V\to P\gamma) \times \frac{\alpha}{3\pi}\frac{1}{q} \sqrt{1-\frac{4m_l^2}{q^2}} (1+\frac{2m^2_l}{q^2}) \\
&\times\left[(1+\frac{q^2}{m_V^2-m_P^2})^2-\frac{4m_V^2q^2}{(m^2_V-m_P^2)^2}\right]^{3/2},
\end{split}
\end{equation}
where $q = M_{e^+e^-}$, $V=J/\psi$, and $P=\eta_c$. Considering the non-negligible width of $\eta_c$, a numerical convolution is performed as
\begin{equation}
\frac{d\Gamma}{d M_{e^+e^-}} \to \int{\frac{d\Gamma}{d M_{e^+e^-}}(m_{\eta_c})\times BW(m_{\eta_c})\times d m_{\eta_c}},
\end{equation}
where $BW(m_{\eta_c})$ is the line shape of $\eta_c$. Finally, the normalized form factors $F^i_{\rm norm} = F^i/F^0$ are determined as summarized in Table~\ref{tab:FF_measurement}. 

Table~\ref{tab:FF_sys} summarizes the systematic uncertainties of $F^i_{\rm norm}$. The total systematic uncertainty is obtained by adding the individual contributions in quadrature.
\begin{itemize}
	\item \textit{MC statistics}: The systematic uncertainty due to MC statistics is evaluated by generating 100 alternative efficiency matrices $\epsilon_{ij}$, where the matrix elements are varied according to their statistical uncertainties. The values of $F^i_{\rm norm}$ are recalculated for each case, and the standard deviation of the resulting distribution is taken as the corresponding uncertainty.
	\item \textit{Fit to $M^{\rm rec}_{e^+e^-}$}: The systematic uncertainty associated with the fit to $M^{\rm rec}_{e^+e^-}$ is estimated in the same manner as for the relative branching fraction measurement. The difference between the nominal and alternative fit results is taken as the corresponding uncertainty.
	\item \textit{$\gamma$-conversion background}: The systematic uncertainty due to the $\gamma$-conversion background is evaluated by generating 100 alternative values of the ratio $r_{\gamma-{\rm conv}}$, varied according to the statistical uncertainties listed in Table~\ref{tab:FF_measurement}. The values of $F^i_{\rm norm}$ are recalculated for each variation, and the standard deviation of the resulting distribution is assigned as the corresponding uncertainty.
\end{itemize}

\begin{table}[htbp]
	\centering
	\tabcolsep=0.22cm
	\caption{Systematic uncertainties of $F^i_{\rm norm}$ in each $M_{e^+e^-}$ interval.}
	\begin{tabular}{ccccc}
		\hline\hline
		$M_{e^+e^-}$~(MeV/$c^2$) & MC stat. & Fit & $\gamma$-conv. & Sum\\
		\hline
		$(10,20)$ & 0.02 & 0.39 & 0.02 & 0.39 \\ 
		$(20,30)$ & 0.04 & 0.33 & 0.04 & 0.33 \\ 
		$(30,150)$ & 0.04 & 0.33 & 0.04 & 0.34 \\ 
		\hline
		\hline
	\end{tabular}
	\label{tab:FF_sys}
\end{table}

The extracted $F^i_{\rm norm}$ values are shown in Fig.~\ref{fig:FFs}. No significant deviation from the VMD prediction or the Lattice-QCD calculation~\cite{Colquhoun:2023zbc} is observed.

\section{SUMMARY}

Using $(10.087\pm0.044)\times10^9$ $J/\psi$ events, the EM Dalitz decay $J/\psi\to e^+e^-\eta_c$ is observed for the first time. The relative branching fraction $R \equiv \frac{\Gamma(J/\psi\to e^+e^-\eta_c)}{\Gamma(J/\psi\to \gamma\eta_c)}$ is determined to be $(0.65\pm0.02_{\rm stat.}\pm0.07_{\rm sys.})\%$, where the first uncertainty is statistical and the second is systematic. The measured value of $R$ is consistent with the prediction of the VMD model, $R_{\rm th}=0.606\%$~\cite{Gell-Mann:1961jim,Bauer:1977iq}, and with the Lattice-QCD calculation, $R_{\rm th}=0.608\%$~\cite{Colquhoun:2023zbc}.

Combining this result with $\mathcal{B}(J/\psi\to \gamma\eta_c)=(1.41\pm0.14)\%$~\cite{ParticleDataGroup:2024cfk}, the absolute branching fraction is determined to be $\mathcal{B}(J/\psi\to e^+e^-\eta_c)=(0.92\pm0.03_{\rm stat.}\pm0.10_{\rm syst.}\pm0.09_{\rm input})\times10^{-4}$, where the third uncertainty arises from the input branching fraction. The $q^2$-dependent form factors are also extracted, and no significant deviation from the VMD prediction or the Lattice-QCD calculation~\cite{Colquhoun:2023zbc} is observed.

\section{ACKNOWLEDGMENTS}

The BESIII Collaboration thanks the staff of BEPCII (https://cstr.cn/31109.02.BEPC) and the IHEP computing center for their strong support. This work is supported in part by National Key R\&D Program of China under Contracts Nos. 2025YFA1613900, 2023YFA1606000, 2023YFA1606704, 2025YFA1613900; National Natural Science Foundation of China (NSFC) under Contracts Nos. 11635010, 11935015, 11935016, 11935018, 12025502, 12035009, 12035013, 12061131003, 12192260, 12192261, 12192262, 12192263, 12192264, 12192265, 12221005, 12225509, 12235017, 12342502, 12361141819, 12535005; the Chinese Academy of Sciences (CAS) Large-Scale Scientific Facility Program; the Strategic Priority Research Program of Chinese Academy of Sciences under Contract No. XDA0480600; CAS under Contract No. YSBR-101; 100 Talents Program of CAS; The Institute of Nuclear and Particle Physics (INPAC) and Shanghai Key Laboratory for Particle Physics and Cosmology; ERC under Contract No. 758462; German Research Foundation DFG under Contract No. FOR5327; Istituto Nazionale di Fisica Nucleare, Italy; Knut and Alice Wallenberg Foundation under Contracts Nos. 2021.0174, 2021.0299, 2023.0315; Ministry of Development of Turkey under Contract No. DPT2006K-120470; National Research Foundation of Korea under Contract No. NRF-2022R1A2C1092335; National Science and Technology fund of Mongolia; Polish National Science Centre under Contract No. 2024/53/B/ST2/00975; STFC (United Kingdom); Swedish Research Council under Contract No. 2019.04595; U. S. Department of Energy under Contract No. DE-FG02-05ER41374

\bibliography{reference}

\begin{thebibliography}{41}%
\makeatletter
\providecommand \@ifxundefined [1]{%
 \@ifx{#1\undefined}
}%
\providecommand \@ifnum [1]{%
 \ifnum #1\expandafter \@firstoftwo
 \else \expandafter \@secondoftwo
 \fi
}%
\providecommand \@ifx [1]{%
 \ifx #1\expandafter \@firstoftwo
 \else \expandafter \@secondoftwo
 \fi
}%
\providecommand \natexlab [1]{#1}%
\providecommand \enquote  [1]{``#1''}%
\providecommand \bibnamefont  [1]{#1}%
\providecommand \bibfnamefont [1]{#1}%
\providecommand \citenamefont [1]{#1}%
\providecommand \href@noop [0]{\@secondoftwo}%
\providecommand \href [0]{\begingroup \@sanitize@url \@href}%
\providecommand \@href[1]{\@@startlink{#1}\@@href}%
\providecommand \@@href[1]{\endgroup#1\@@endlink}%
\providecommand \@sanitize@url [0]{\catcode `\\12\catcode `\$12\catcode
  `\&12\catcode `\#12\catcode `\^12\catcode `\_12\catcode `\%12\relax}%
\providecommand \@@startlink[1]{}%
\providecommand \@@endlink[0]{}%
\providecommand \url  [0]{\begingroup\@sanitize@url \@url }%
\providecommand \@url [1]{\endgroup\@href {#1}{\urlprefix }}%
\providecommand \urlprefix  [0]{URL }%
\providecommand \Eprint [0]{\href }%
\providecommand \doibase [0]{http://dx.doi.org/}%
\providecommand \selectlanguage [0]{\@gobble}%
\providecommand \bibinfo  [0]{\@secondoftwo}%
\providecommand \bibfield  [0]{\@secondoftwo}%
\providecommand \translation [1]{[#1]}%
\providecommand \BibitemOpen [0]{}%
\providecommand \bibitemStop [0]{}%
\providecommand \bibitemNoStop [0]{.\EOS\space}%
\providecommand \EOS [0]{\spacefactor3000\relax}%
\providecommand \BibitemShut  [1]{\csname bibitem#1\endcsname}%
\let\auto@bib@innerbib\@empty
\bibitem [{\citenamefont {Mitchell}\ \emph {et~al.}(2009)\citenamefont
  {Mitchell} \emph {et~al.}}]{CLEO:2008pln}%
  \BibitemOpen
  \bibfield  {author} {\bibinfo {author} {\bibfnamefont {R.~E.}\ \bibnamefont
  {Mitchell}} \emph {et~al.} (\bibinfo {collaboration} {CLEO Collaboration}),\
  }\href {\doibase 10.1103/PhysRevLett.102.011801} {\bibfield  {journal}
  {\bibinfo  {journal} {Phys. Rev. Lett.}\ }\textbf {\bibinfo {volume} {102}},\
  \bibinfo {pages} {011801} (\bibinfo {year} {2009})},\ \bibinfo {note}
  {[Erratum: Phys.Rev.Lett. 106, 159903 (2011)]}\BibitemShut {NoStop}%
\bibitem [{\citenamefont {Anashin}\ \emph {et~al.}(2010)\citenamefont {Anashin}
  \emph {et~al.}}]{Anashin:2010nr}%
  \BibitemOpen
  \bibfield  {author} {\bibinfo {author} {\bibfnamefont {V.~V.}\ \bibnamefont
  {Anashin}} \emph {et~al.},\ }\href {\doibase 10.1088/1674-1137/34/6/035}
  {\bibfield  {journal} {\bibinfo  {journal} {Chin. Phys. C}\ }\textbf
  {\bibinfo {volume} {34}},\ \bibinfo {pages} {831} (\bibinfo {year}
  {2010})}\BibitemShut {NoStop}%
\bibitem [{\citenamefont {Anashin}\ \emph {et~al.}(2014)\citenamefont {Anashin}
  \emph {et~al.}}]{Anashin:2014wva}%
  \BibitemOpen
  \bibfield  {author} {\bibinfo {author} {\bibfnamefont {V.~V.}\ \bibnamefont
  {Anashin}} \emph {et~al.},\ }\href {\doibase 10.1016/j.physletb.2014.09.064}
  {\bibfield  {journal} {\bibinfo  {journal} {Phys. Lett. B}\ }\textbf
  {\bibinfo {volume} {738}},\ \bibinfo {pages} {391} (\bibinfo {year}
  {2014})}\BibitemShut {NoStop}%
\bibitem [{\citenamefont {Beilin}\ and\ \citenamefont
  {Radyushkin}(1985)}]{Beilin:1984pf}%
  \BibitemOpen
  \bibfield  {author} {\bibinfo {author} {\bibfnamefont {V.~A.}\ \bibnamefont
  {Beilin}}\ and\ \bibinfo {author} {\bibfnamefont {A.~V.}\ \bibnamefont
  {Radyushkin}},\ }\href {\doibase 10.1016/0550-3213(85)90310-4} {\bibfield
  {journal} {\bibinfo  {journal} {Nucl. Phys. B}\ }\textbf {\bibinfo {volume}
  {260}},\ \bibinfo {pages} {61} (\bibinfo {year} {1985})}\BibitemShut
  {NoStop}%
\bibitem [{\citenamefont {Brambilla}\ \emph {et~al.}(2006)\citenamefont
  {Brambilla}, \citenamefont {Jia},\ and\ \citenamefont
  {Vairo}}]{Brambilla:2005zw}%
  \BibitemOpen
  \bibfield  {author} {\bibinfo {author} {\bibfnamefont {N.}~\bibnamefont
  {Brambilla}}, \bibinfo {author} {\bibfnamefont {Y.}~\bibnamefont {Jia}}, \
  and\ \bibinfo {author} {\bibfnamefont {A.}~\bibnamefont {Vairo}},\ }\href
  {\doibase 10.1103/PhysRevD.73.054005} {\bibfield  {journal} {\bibinfo
  {journal} {Phys. Rev. D}\ }\textbf {\bibinfo {volume} {73}},\ \bibinfo
  {pages} {054005} (\bibinfo {year} {2006})}\BibitemShut {NoStop}%
\bibitem [{\citenamefont {Pineda}\ and\ \citenamefont
  {Segovia}(2013)}]{Pineda:2013lta}%
  \BibitemOpen
  \bibfield  {author} {\bibinfo {author} {\bibfnamefont {A.}~\bibnamefont
  {Pineda}}\ and\ \bibinfo {author} {\bibfnamefont {J.}~\bibnamefont
  {Segovia}},\ }\href {\doibase 10.1103/PhysRevD.87.074024} {\bibfield
  {journal} {\bibinfo  {journal} {Phys. Rev. D}\ }\textbf {\bibinfo {volume}
  {87}},\ \bibinfo {pages} {074024} (\bibinfo {year} {2013})}\BibitemShut
  {NoStop}%
\bibitem [{\citenamefont {Abe}\ \emph {et~al.}(2002)\citenamefont {Abe} \emph
  {et~al.}}]{Belle:2002tfa}%
  \BibitemOpen
  \bibfield  {author} {\bibinfo {author} {\bibfnamefont {K.}~\bibnamefont
  {Abe}} \emph {et~al.} (\bibinfo {collaboration} {Belle Collaboration}),\
  }\href {\doibase 10.1103/PhysRevLett.89.142001} {\bibfield  {journal}
  {\bibinfo  {journal} {Phys. Rev. Lett.}\ }\textbf {\bibinfo {volume} {89}},\
  \bibinfo {pages} {142001} (\bibinfo {year} {2002})}\BibitemShut {NoStop}%
\bibitem [{\citenamefont {Abe}\ \emph {et~al.}(2004)\citenamefont {Abe} \emph
  {et~al.}}]{Belle:2004abn}%
  \BibitemOpen
  \bibfield  {author} {\bibinfo {author} {\bibfnamefont {K.}~\bibnamefont
  {Abe}} \emph {et~al.} (\bibinfo {collaboration} {Belle Collaboration}),\
  }\href {\doibase 10.1103/PhysRevD.70.071102} {\bibfield  {journal} {\bibinfo
  {journal} {Phys. Rev. D}\ }\textbf {\bibinfo {volume} {70}},\ \bibinfo
  {pages} {071102} (\bibinfo {year} {2004})}\BibitemShut {NoStop}%
\bibitem [{\citenamefont {Aubert}\ \emph {et~al.}(2005)\citenamefont {Aubert}
  \emph {et~al.}}]{BaBar:2005nic}%
  \BibitemOpen
  \bibfield  {author} {\bibinfo {author} {\bibfnamefont {B.}~\bibnamefont
  {Aubert}} \emph {et~al.} (\bibinfo {collaboration} {BaBar Collaboration}),\
  }\href {\doibase 10.1103/PhysRevD.72.031101} {\bibfield  {journal} {\bibinfo
  {journal} {Phys. Rev. D}\ }\textbf {\bibinfo {volume} {72}},\ \bibinfo
  {pages} {031101} (\bibinfo {year} {2005})}\BibitemShut {NoStop}%
\bibitem [{\citenamefont {Huang}\ \emph {et~al.}(2023)\citenamefont {Huang},
  \citenamefont {Gong},\ and\ \citenamefont {Wang}}]{Huang:2022dfw}%
  \BibitemOpen
  \bibfield  {author} {\bibinfo {author} {\bibfnamefont {X.~D.}\ \bibnamefont
  {Huang}}, \bibinfo {author} {\bibfnamefont {B.}~\bibnamefont {Gong}}, \ and\
  \bibinfo {author} {\bibfnamefont {J.~X.}\ \bibnamefont {Wang}},\ }\href
  {\doibase 10.1007/JHEP02(2023)049} {\bibfield  {journal} {\bibinfo  {journal}
  {JHEP}\ }\textbf {\bibinfo {volume} {02}},\ \bibinfo {pages} {049} (\bibinfo
  {year} {2023})}\BibitemShut {NoStop}%
\bibitem [{\citenamefont {Sun}(2021)}]{Sun:2021tma}%
  \BibitemOpen
  \bibfield  {author} {\bibinfo {author} {\bibfnamefont {Z.}~\bibnamefont
  {Sun}},\ }\href {\doibase 10.1007/JHEP09(2021)073} {\bibfield  {journal}
  {\bibinfo  {journal} {JHEP}\ }\textbf {\bibinfo {volume} {09}},\ \bibinfo
  {pages} {073} (\bibinfo {year} {2021})}\BibitemShut {NoStop}%
\bibitem [{\citenamefont {Zeng}\ \emph {et~al.}(2021)\citenamefont {Zeng},
  \citenamefont {Fu}, \citenamefont {Hu}, \citenamefont {Chen}, \citenamefont
  {Cheng},\ and\ \citenamefont {Wu}}]{Zeng:2021hwt}%
  \BibitemOpen
  \bibfield  {author} {\bibinfo {author} {\bibfnamefont {L.}~\bibnamefont
  {Zeng}}, \bibinfo {author} {\bibfnamefont {H.~B.}\ \bibnamefont {Fu}},
  \bibinfo {author} {\bibfnamefont {D.~D.}\ \bibnamefont {Hu}}, \bibinfo
  {author} {\bibfnamefont {L.~L.}\ \bibnamefont {Chen}}, \bibinfo {author}
  {\bibfnamefont {W.}~\bibnamefont {Cheng}}, \ and\ \bibinfo {author}
  {\bibfnamefont {X.~G.}\ \bibnamefont {Wu}},\ }\href {\doibase
  10.1103/PhysRevD.103.056012} {\bibfield  {journal} {\bibinfo  {journal}
  {Phys. Rev. D}\ }\textbf {\bibinfo {volume} {103}},\ \bibinfo {pages}
  {056012} (\bibinfo {year} {2021})}\BibitemShut {NoStop}%
\bibitem [{\citenamefont {Ablikim}\ \emph
  {et~al.}(2017{\natexlab{a}})\citenamefont {Ablikim} \emph
  {et~al.}}]{BESIII:2017ung}%
  \BibitemOpen
  \bibfield  {author} {\bibinfo {author} {\bibfnamefont {M.}~\bibnamefont
  {Ablikim}} \emph {et~al.} (\bibinfo {collaboration} {BESIII Collaboration}),\
  }\href {\doibase 10.1103/PhysRevLett.118.221802} {\bibfield  {journal}
  {\bibinfo  {journal} {Phys. Rev. Lett.}\ }\textbf {\bibinfo {volume} {118}},\
  \bibinfo {pages} {221802} (\bibinfo {year} {2017}{\natexlab{a}})}\BibitemShut
  {NoStop}%
\bibitem [{\citenamefont {Ablikim}\ \emph
  {et~al.}(2022{\natexlab{a}})\citenamefont {Ablikim} \emph
  {et~al.}}]{BESIII:2022jde}%
  \BibitemOpen
  \bibfield  {author} {\bibinfo {author} {\bibfnamefont {M.}~\bibnamefont
  {Ablikim}} \emph {et~al.} (\bibinfo {collaboration} {BESIII Collaboration}),\
  }\href {\doibase 10.1103/PhysRevD.106.112002} {\bibfield  {journal} {\bibinfo
   {journal} {Phys. Rev. D}\ }\textbf {\bibinfo {volume} {106}},\ \bibinfo
  {pages} {112002} (\bibinfo {year} {2022}{\natexlab{a}})}\BibitemShut
  {NoStop}%
\bibitem [{\citenamefont {Ablikim}\ \emph
  {et~al.}(2019{\natexlab{a}})\citenamefont {Ablikim} \emph
  {et~al.}}]{BESIII:2019yeu}%
  \BibitemOpen
  \bibfield  {author} {\bibinfo {author} {\bibfnamefont {M.}~\bibnamefont
  {Ablikim}} \emph {et~al.} (\bibinfo {collaboration} {BESIII Collaboration}),\
  }\href {\doibase 10.1103/PhysRevD.99.051101} {\bibfield  {journal} {\bibinfo
  {journal} {Phys. Rev. D}\ }\textbf {\bibinfo {volume} {99}},\ \bibinfo
  {pages} {051101} (\bibinfo {year} {2019}{\natexlab{a}})}\BibitemShut
  {NoStop}%
\bibitem [{\citenamefont {Ablikim}\ \emph {et~al.}(2024)\citenamefont {Ablikim}
  \emph {et~al.}}]{BESIII:2024kkf}%
  \BibitemOpen
  \bibfield  {author} {\bibinfo {author} {\bibfnamefont {M.}~\bibnamefont
  {Ablikim}} \emph {et~al.} (\bibinfo {collaboration} {BESIII Collaboration}),\
  }\href {\doibase 10.1103/PhysRevD.110.L111101} {\bibfield  {journal}
  {\bibinfo  {journal} {Phys. Rev. D}\ }\textbf {\bibinfo {volume} {110}},\
  \bibinfo {pages} {L111101} (\bibinfo {year} {2024})}\BibitemShut {NoStop}%
\bibitem [{\citenamefont {Ablikim}\ \emph
  {et~al.}(2022{\natexlab{b}})\citenamefont {Ablikim} \emph
  {et~al.}}]{BESIII:2021cxx}%
  \BibitemOpen
  \bibfield  {author} {\bibinfo {author} {\bibfnamefont {M.}~\bibnamefont
  {Ablikim}} \emph {et~al.} (\bibinfo {collaboration} {BESIII Collaboration}),\
  }\href {\doibase 10.1088/1674-1137/ac5c2e} {\bibfield  {journal} {\bibinfo
  {journal} {Chin. Phys. C}\ }\textbf {\bibinfo {volume} {46}},\ \bibinfo
  {pages} {074001} (\bibinfo {year} {2022}{\natexlab{b}})}\BibitemShut
  {NoStop}%
\bibitem [{\citenamefont {Gell~Mann}\ and\ \citenamefont
  {Zachariasen}(1961)}]{Gell-Mann:1961jim}%
  \BibitemOpen
  \bibfield  {author} {\bibinfo {author} {\bibfnamefont {M.}~\bibnamefont
  {Gell~Mann}}\ and\ \bibinfo {author} {\bibfnamefont {F.}~\bibnamefont
  {Zachariasen}},\ }\href {\doibase 10.1103/PhysRev.124.953} {\bibfield
  {journal} {\bibinfo  {journal} {Phys. Rev.}\ }\textbf {\bibinfo {volume}
  {124}},\ \bibinfo {pages} {953} (\bibinfo {year} {1961})}\BibitemShut
  {NoStop}%
\bibitem [{\citenamefont {Bauer}\ \emph {et~al.}(1978)\citenamefont {Bauer},
  \citenamefont {Spital}, \citenamefont {Yennie},\ and\ \citenamefont
  {Pipkin}}]{Bauer:1977iq}%
  \BibitemOpen
  \bibfield  {author} {\bibinfo {author} {\bibfnamefont {T.~H.}\ \bibnamefont
  {Bauer}}, \bibinfo {author} {\bibfnamefont {R.~D.}\ \bibnamefont {Spital}},
  \bibinfo {author} {\bibfnamefont {D.~R.}\ \bibnamefont {Yennie}}, \ and\
  \bibinfo {author} {\bibfnamefont {F.~M.}\ \bibnamefont {Pipkin}},\ }\href
  {\doibase 10.1103/RevModPhys.50.261} {\bibfield  {journal} {\bibinfo
  {journal} {Rev. Mod. Phys.}\ }\textbf {\bibinfo {volume} {50}},\ \bibinfo
  {pages} {261} (\bibinfo {year} {1978})},\ \bibinfo {note} {[Erratum:
  Rev.Mod.Phys. 51, 407 (1979)]}\BibitemShut {NoStop}%
\bibitem [{\citenamefont {Colquhoun}\ \emph {et~al.}(2023)\citenamefont
  {Colquhoun}, \citenamefont {Cooper}, \citenamefont {Davies},\ and\
  \citenamefont {Lepage}}]{Colquhoun:2023zbc}%
  \BibitemOpen
  \bibfield  {author} {\bibinfo {author} {\bibfnamefont {B.}~\bibnamefont
  {Colquhoun}}, \bibinfo {author} {\bibfnamefont {L.~J.}\ \bibnamefont
  {Cooper}}, \bibinfo {author} {\bibfnamefont {C.~T.~H.}\ \bibnamefont
  {Davies}}, \ and\ \bibinfo {author} {\bibfnamefont {G.~P.}\ \bibnamefont
  {Lepage}} (\bibinfo {collaboration} {Particle Data Group and HPQCD
  Collaboration}),\ }\href {\doibase 10.1103/PhysRevD.108.014513} {\bibfield
  {journal} {\bibinfo  {journal} {Phys. Rev. D}\ }\textbf {\bibinfo {volume}
  {108}},\ \bibinfo {pages} {014513} (\bibinfo {year} {2023})}\BibitemShut
  {NoStop}%
\bibitem [{\citenamefont {Ablikim}\ \emph {et~al.}(2010)\citenamefont {Ablikim}
  \emph {et~al.}}]{BESIII:2009fln}%
  \BibitemOpen
  \bibfield  {author} {\bibinfo {author} {\bibfnamefont {M.}~\bibnamefont
  {Ablikim}} \emph {et~al.} (\bibinfo {collaboration} {BESIII Collaboration}),\
  }\href {\doibase 10.1016/j.nima.2009.12.050} {\bibfield  {journal} {\bibinfo
  {journal} {Nucl. Instrum. Meth. A}\ }\textbf {\bibinfo {volume} {614}},\
  \bibinfo {pages} {345} (\bibinfo {year} {2010})}\BibitemShut {NoStop}%
\bibitem [{\citenamefont {Yu}\ \emph {et~al.}(2016)\citenamefont {Yu} \emph
  {et~al.}}]{Yu:2016cof}%
  \BibitemOpen
  \bibfield  {author} {\bibinfo {author} {\bibfnamefont {C.}~\bibnamefont {Yu}}
  \emph {et~al.},\ }in\ \href {\doibase 10.18429/JACoW-IPAC2016-TUYA01} {\emph
  {\bibinfo {booktitle} {{7th International Particle Accelerator
  Conference}}}}\ (\bibinfo {year} {2016})\ p.\ \bibinfo {pages}
  {TUYA01}\BibitemShut {NoStop}%
\bibitem [{\citenamefont {Ablikim}\ \emph {et~al.}(2020)\citenamefont {Ablikim}
  \emph {et~al.}}]{BESIII:2020nme}%
  \BibitemOpen
  \bibfield  {author} {\bibinfo {author} {\bibfnamefont {M.}~\bibnamefont
  {Ablikim}} \emph {et~al.} (\bibinfo {collaboration} {BESIII Collaboration}),\
  }\href {\doibase 10.1088/1674-1137/44/4/040001} {\bibfield  {journal}
  {\bibinfo  {journal} {Chin. Phys. C}\ }\textbf {\bibinfo {volume} {44}},\
  \bibinfo {pages} {040001} (\bibinfo {year} {2020})}\BibitemShut {NoStop}%
\bibitem [{\citenamefont {Lu}\ \emph {et~al.}(2020)\citenamefont {Lu},
  \citenamefont {Xiao},\ and\ \citenamefont {Ji}}]{Lu:2020imt}%
  \BibitemOpen
  \bibfield  {author} {\bibinfo {author} {\bibfnamefont {J.}~\bibnamefont
  {Lu}}, \bibinfo {author} {\bibfnamefont {Y.}~\bibnamefont {Xiao}}, \ and\
  \bibinfo {author} {\bibfnamefont {X.}~\bibnamefont {Ji}},\ }\href {\doibase
  10.1007/s41605-020-00188-8} {\bibfield  {journal} {\bibinfo  {journal}
  {Radiat. Detect. Technol. Methods}\ }\textbf {\bibinfo {volume} {4}},\
  \bibinfo {pages} {337} (\bibinfo {year} {2020})}\BibitemShut {NoStop}%
\bibitem [{\citenamefont {Zhang}\ \emph {et~al.}(2022)\citenamefont {Zhang}
  \emph {et~al.}}]{Zhang:2022bdc}%
  \BibitemOpen
  \bibfield  {author} {\bibinfo {author} {\bibfnamefont {J.~W.}\ \bibnamefont
  {Zhang}} \emph {et~al.},\ }\href {\doibase 10.1007/s41605-022-00331-7}
  {\bibfield  {journal} {\bibinfo  {journal} {Radiat. Detect. Technol.
  Methods}\ }\textbf {\bibinfo {volume} {6}},\ \bibinfo {pages} {289} (\bibinfo
  {year} {2022})}\BibitemShut {NoStop}%
\bibitem [{\citenamefont {Li}\ \emph {et~al.}(2017)\citenamefont {Li} \emph
  {et~al.}}]{Li:2017jpg}%
  \BibitemOpen
  \bibfield  {author} {\bibinfo {author} {\bibfnamefont {X.}~\bibnamefont {Li}}
  \emph {et~al.},\ }\href {\doibase 10.1007/s41605-017-0014-2} {\bibfield
  {journal} {\bibinfo  {journal} {Radiat. Detect. Technol. Methods}\ }\textbf
  {\bibinfo {volume} {1}},\ \bibinfo {pages} {13} (\bibinfo {year}
  {2017})}\BibitemShut {NoStop}%
\bibitem [{\citenamefont {Guo}\ \emph {et~al.}(2017)\citenamefont {Guo} \emph
  {et~al.}}]{Guo:2017sjt}%
  \BibitemOpen
  \bibfield  {author} {\bibinfo {author} {\bibfnamefont {Y.~X.}\ \bibnamefont
  {Guo}} \emph {et~al.},\ }\href {\doibase 10.1007/s41605-017-0012-4}
  {\bibfield  {journal} {\bibinfo  {journal} {Radiat. Detect. Technol.
  Methods}\ }\textbf {\bibinfo {volume} {1}},\ \bibinfo {pages} {15} (\bibinfo
  {year} {2017})}\BibitemShut {NoStop}%
\bibitem [{\citenamefont {Cao}\ \emph {et~al.}(2020)\citenamefont {Cao} \emph
  {et~al.}}]{Cao:2020ibk}%
  \BibitemOpen
  \bibfield  {author} {\bibinfo {author} {\bibfnamefont {P.}~\bibnamefont
  {Cao}} \emph {et~al.},\ }\href {\doibase 10.1016/j.nima.2019.163053}
  {\bibfield  {journal} {\bibinfo  {journal} {Nucl. Instrum. Meth. A}\ }\textbf
  {\bibinfo {volume} {953}},\ \bibinfo {pages} {163053} (\bibinfo {year}
  {2020})}\BibitemShut {NoStop}%
\bibitem [{\citenamefont {Agostinelli}\ \emph {et~al.}(2003)\citenamefont
  {Agostinelli} \emph {et~al.}}]{GEANT4:2002zbu}%
  \BibitemOpen
  \bibfield  {author} {\bibinfo {author} {\bibfnamefont {S.}~\bibnamefont
  {Agostinelli}} \emph {et~al.} (\bibinfo {collaboration} {GEANT4
  Collaboration}),\ }\href {\doibase 10.1016/S0168-9002(03)01368-8} {\bibfield
  {journal} {\bibinfo  {journal} {Nucl. Instrum. Meth. A}\ }\textbf {\bibinfo
  {volume} {506}},\ \bibinfo {pages} {250} (\bibinfo {year}
  {2003})}\BibitemShut {NoStop}%
\bibitem [{\citenamefont {Jadach}\ \emph {et~al.}(2000)\citenamefont {Jadach},
  \citenamefont {Ward},\ and\ \citenamefont {Was}}]{Jadach:1999vf}%
  \BibitemOpen
  \bibfield  {author} {\bibinfo {author} {\bibfnamefont {S.}~\bibnamefont
  {Jadach}}, \bibinfo {author} {\bibfnamefont {B.~F.~L.}\ \bibnamefont {Ward}},
  \ and\ \bibinfo {author} {\bibfnamefont {Z.}~\bibnamefont {Was}},\ }\href
  {\doibase 10.1016/S0010-4655(00)00048-5} {\bibfield  {journal} {\bibinfo
  {journal} {Comput. Phys. Commun.}\ }\textbf {\bibinfo {volume} {130}},\
  \bibinfo {pages} {260} (\bibinfo {year} {2000})}\BibitemShut {NoStop}%
\bibitem [{\citenamefont {Jadach}\ \emph {et~al.}(2001)\citenamefont {Jadach},
  \citenamefont {Ward},\ and\ \citenamefont {Was}}]{Jadach:2000ir}%
  \BibitemOpen
  \bibfield  {author} {\bibinfo {author} {\bibfnamefont {S.}~\bibnamefont
  {Jadach}}, \bibinfo {author} {\bibfnamefont {B.~F.~L.}\ \bibnamefont {Ward}},
  \ and\ \bibinfo {author} {\bibfnamefont {Z.}~\bibnamefont {Was}},\ }\href
  {\doibase 10.1103/PhysRevD.63.113009} {\bibfield  {journal} {\bibinfo
  {journal} {Phys. Rev. D}\ }\textbf {\bibinfo {volume} {63}},\ \bibinfo
  {pages} {113009} (\bibinfo {year} {2001})}\BibitemShut {NoStop}%
\bibitem [{\citenamefont {Lange}(2001)}]{Lange:2001uf}%
  \BibitemOpen
  \bibfield  {author} {\bibinfo {author} {\bibfnamefont {D.~J.}\ \bibnamefont
  {Lange}},\ }\href {\doibase 10.1016/S0168-9002(01)00089-4} {\bibfield
  {journal} {\bibinfo  {journal} {Nucl. Instrum. Meth. A}\ }\textbf {\bibinfo
  {volume} {462}},\ \bibinfo {pages} {152} (\bibinfo {year}
  {2001})}\BibitemShut {NoStop}%
\bibitem [{\citenamefont {Ping}(2008)}]{Ping:2008zz}%
  \BibitemOpen
  \bibfield  {author} {\bibinfo {author} {\bibfnamefont {R.-G.}\ \bibnamefont
  {Ping}},\ }\href {\doibase 10.1088/1674-1137/32/8/001} {\bibfield  {journal}
  {\bibinfo  {journal} {Chin. Phys. C}\ }\textbf {\bibinfo {volume} {32}},\
  \bibinfo {pages} {599} (\bibinfo {year} {2008})}\BibitemShut {NoStop}%
\bibitem [{\citenamefont {Navas}\ \emph {et~al.}(2024)\citenamefont {Navas}
  \emph {et~al.}}]{ParticleDataGroup:2024cfk}%
  \BibitemOpen
  \bibfield  {author} {\bibinfo {author} {\bibfnamefont {S.}~\bibnamefont
  {Navas}} \emph {et~al.} (\bibinfo {collaboration} {Particle Data Group}),\
  }\href {\doibase 10.1103/PhysRevD.110.030001} {\bibfield  {journal} {\bibinfo
   {journal} {Phys. Rev. D}\ }\textbf {\bibinfo {volume} {110}},\ \bibinfo
  {pages} {030001} (\bibinfo {year} {2024})}\BibitemShut {NoStop}%
\bibitem [{\citenamefont {Chen}\ \emph {et~al.}(2000)\citenamefont {Chen},
  \citenamefont {Huang}, \citenamefont {Qi}, \citenamefont {Zhang},\ and\
  \citenamefont {Zhu}}]{Chen:2000tv}%
  \BibitemOpen
  \bibfield  {author} {\bibinfo {author} {\bibfnamefont {J.~C.}\ \bibnamefont
  {Chen}}, \bibinfo {author} {\bibfnamefont {G.~S.}\ \bibnamefont {Huang}},
  \bibinfo {author} {\bibfnamefont {X.~R.}\ \bibnamefont {Qi}}, \bibinfo
  {author} {\bibfnamefont {D.~H.}\ \bibnamefont {Zhang}}, \ and\ \bibinfo
  {author} {\bibfnamefont {Y.~S.}\ \bibnamefont {Zhu}},\ }\href {\doibase
  10.1103/PhysRevD.62.034003} {\bibfield  {journal} {\bibinfo  {journal} {Phys.
  Rev. D}\ }\textbf {\bibinfo {volume} {62}},\ \bibinfo {pages} {034003}
  (\bibinfo {year} {2000})}\BibitemShut {NoStop}%
\bibitem [{\citenamefont {Yang}\ \emph {et~al.}(2014)\citenamefont {Yang},
  \citenamefont {Ping},\ and\ \citenamefont {Chen}}]{Yang:2014vra}%
  \BibitemOpen
  \bibfield  {author} {\bibinfo {author} {\bibfnamefont {R.~L.}\ \bibnamefont
  {Yang}}, \bibinfo {author} {\bibfnamefont {R.~G.}\ \bibnamefont {Ping}}, \
  and\ \bibinfo {author} {\bibfnamefont {H.}~\bibnamefont {Chen}},\ }\href
  {\doibase 10.1088/0256-307X/31/6/061301} {\bibfield  {journal} {\bibinfo
  {journal} {Chin. Phys. Lett.}\ }\textbf {\bibinfo {volume} {31}},\ \bibinfo
  {pages} {061301} (\bibinfo {year} {2014})}\BibitemShut {NoStop}%
\bibitem [{\citenamefont {Barberio}\ \emph {et~al.}(1991)\citenamefont
  {Barberio}, \citenamefont {van Eijk},\ and\ \citenamefont
  {Was}}]{Barberio:1990ms}%
  \BibitemOpen
  \bibfield  {author} {\bibinfo {author} {\bibfnamefont {E.}~\bibnamefont
  {Barberio}}, \bibinfo {author} {\bibfnamefont {B.}~\bibnamefont {van Eijk}},
  \ and\ \bibinfo {author} {\bibfnamefont {Z.}~\bibnamefont {Was}},\ }\href
  {\doibase 10.1016/0010-4655(91)90012-A} {\bibfield  {journal} {\bibinfo
  {journal} {Comput. Phys. Commun.}\ }\textbf {\bibinfo {volume} {66}},\
  \bibinfo {pages} {115} (\bibinfo {year} {1991})}\BibitemShut {NoStop}%
\bibitem [{\citenamefont {Gu}\ \emph {et~al.}(2019)\citenamefont {Gu},
  \citenamefont {Li}, \citenamefont {Ma},\ and\ \citenamefont
  {Yang}}]{Gu:2019qwo}%
  \BibitemOpen
  \bibfield  {author} {\bibinfo {author} {\bibfnamefont {L.~M.}\ \bibnamefont
  {Gu}}, \bibinfo {author} {\bibfnamefont {H.~B.}\ \bibnamefont {Li}}, \bibinfo
  {author} {\bibfnamefont {X.~X.}\ \bibnamefont {Ma}}, \ and\ \bibinfo {author}
  {\bibfnamefont {M.~Z.}\ \bibnamefont {Yang}},\ }\href {\doibase
  10.1103/PhysRevD.100.016018} {\bibfield  {journal} {\bibinfo  {journal}
  {Phys. Rev. D}\ }\textbf {\bibinfo {volume} {100}},\ \bibinfo {pages}
  {016018} (\bibinfo {year} {2019})}\BibitemShut {NoStop}%
\bibitem [{\citenamefont {Ablikim}\ \emph
  {et~al.}(2019{\natexlab{b}})\citenamefont {Ablikim} \emph
  {et~al.}}]{BESIII:2018aao}%
  \BibitemOpen
  \bibfield  {author} {\bibinfo {author} {\bibfnamefont {M.}~\bibnamefont
  {Ablikim}} \emph {et~al.} (\bibinfo {collaboration} {BESIII Collaboration}),\
  }\href {\doibase 10.1103/PhysRevD.99.012013} {\bibfield  {journal} {\bibinfo
  {journal} {Phys. Rev. D}\ }\textbf {\bibinfo {volume} {99}},\ \bibinfo
  {pages} {012013} (\bibinfo {year} {2019}{\natexlab{b}})}\BibitemShut
  {NoStop}%
\bibitem [{\citenamefont {Ablikim}\ \emph
  {et~al.}(2017{\natexlab{b}})\citenamefont {Ablikim} \emph
  {et~al.}}]{BESIII:2016kpv}%
  \BibitemOpen
  \bibfield  {author} {\bibinfo {author} {\bibfnamefont {M.}~\bibnamefont
  {Ablikim}} \emph {et~al.} (\bibinfo {collaboration} {BESIII Collaboration}),\
  }\href {\doibase 10.1088/1674-1137/41/1/013001} {\bibfield  {journal}
  {\bibinfo  {journal} {Chin. Phys. C}\ }\textbf {\bibinfo {volume} {41}},\
  \bibinfo {pages} {013001} (\bibinfo {year} {2017}{\natexlab{b}})}\BibitemShut
  {NoStop}%
\bibitem [{\citenamefont {Ablikim}\ \emph {et~al.}(2013)\citenamefont {Ablikim}
  \emph {et~al.}}]{BESIII:2012mpj}%
  \BibitemOpen
  \bibfield  {author} {\bibinfo {author} {\bibfnamefont {M.}~\bibnamefont
  {Ablikim}} \emph {et~al.} (\bibinfo {collaboration} {BESIII Collaboration}),\
  }\href {\doibase 10.1103/PhysRevD.87.012002} {\bibfield  {journal} {\bibinfo
  {journal} {Phys. Rev. D}\ }\textbf {\bibinfo {volume} {87}},\ \bibinfo
  {pages} {012002} (\bibinfo {year} {2013})}\BibitemShut {NoStop}%
\end{thebibliography}%

\end{document}